\documentclass[12pt]{iopart}
\usepackage{graphicx}

\begin{document}

\title[Dynamical correlations in multiorbital Hubbard models:  FLEX]
{Dynamical correlations in multiorbital Hubbard
models:  Fluctuation-exchange approximations}

\author{V.~Drchal *\footnote[3]{E-mail: drchal@fzu.cz} ,
V.~Jani\v{s} *,   J.~Kudrnovsk\'y* , V.~S.~Oudovenko$\sharp$\dag,
X.~Dai\dag, K.~Haule\dag\  and G. Kotliar\dag }
\address{*\ Institute of Physics, AS CR, Na Slovance 2, CZ-182 21 Praha 8,  Czech Republic}%
\address{$\sharp$\ Bogoliubov Laboratory for Theoretical Physics, Joint
Institute for Nuclear Research, 141980 Dubna, Russia}
\address{\dag\ Center for Materials Theory, Department of Physics
and Astronomy, Rutgers University, Piscataway, NJ 08854}

\begin{abstract}
We study the two  band  degenerate Hubbard model  using the
Fluctuation Exchange approximation  (FLEX) and compare the
results with quantum Monte Carlo (QMC) calculations. Both the
self-consistent and the non-self-consistent versions of the FLEX
scheme are investigated. We find that, contrary to the one-band
case, in the multiband case,  good agreement with the quantum
Monte Carlo results is obtained within the electron-electron
$T$-matrix approximation using the full renormalization of the
one-particle propagators. The crossover to strong coupling and
the formation of satellites is more clearly visible in the
non-self-consistent scheme. Finally we discuss the behavior of
the FLEX for higher orbital degeneracy.
\end{abstract}


\maketitle

\section{Introduction}
\label{sec:intro}

Recent progress in the development and application of advanced
experimental techniques made available a number of new materials
and compounds with strongly correlated electrons that cannot be
entirely described within the Density Functional Theory (DFT)
\cite{HK,WK}. Some transition-metal alloys, cuprates, manganites,
heavy-fermion Kondo and mixed-valence systems, as well as
lanthanides and transuranium compounds behave in a way that can
neither be explained nor understood within first-principles
computational schemes based on local approximations of the
correlation-exchange potential in the DFT. It is necessary to
take explicitly into account dynamical fluctuations to describe
these materials. It can be done at best and in a manageable way
within the Dynamical Mean Field Theory (DMFT) \cite{GKKR}. The
DMFT captures most of the relevant local quantum dynamical
effects of electron correlations. Recent progress in the
development of various impurity solvers for the DMFT has opened a
way for combining advanced many-body techniques with ab initio
methods to build realistic computational schemes for materials
with strongly correlated electrons.

The standard way of combining ab initio calculations with the
DMFT is via multiorbital Hubbard models. The input parameters for
multiorbital Hubbard models to be solved within the DMFT are
determined from ab initio calculations, for example using the
linearized muffin-tin orbital (LMTO) method~\cite{TDKSW}. No
universal scheme to solve the DMFT equations exactly does exist.
Hence, one has to resort to approximate impurity solvers to reach
quantitative results from the DMFT. We distinguish essentially
two types of DMFT solvers: numerical schemes and analytical
methods based on many-body perturbation theory. The former
schemes aim at numerically exact quantitative solutions while the
latter at analytically controllable schemes. Although analytic
methods are not quantitatively as accurate as the numerical
solutions, they have an appealing feature in that they offer an
analytically controllable approach with a direct access to
spectral functions on the real frequency axis. Analytic
approaches are needed in most situations to complement the
numerical solutions so that we can assess the peak structure of
spectral functions when performing analytic continuation of
numerical results from the imaginary axis of Matsubara
frequencies. To gain confidence in approaches based on many-body
perturbation theory we should test them in simpler situations and
compare their results with available more precise numerical
simulations.

In the context of the one-band model, extensive comparisons
between perturbative approaches such as the the Iterated
Perturbation Theory (IPT) and the QMC method have been carried
out (see, for example, Refs.~\cite{Kajueter:1996,GKKR}). It
appeared that the non-self-consistent IPT is a rather accurate
approximation. Analytic extension of the IPT and second-order
perturbation theory via multiple two-particle scatterings, the
FLEX, has already been applied to iron and nickel in
self-consistent and non-self-consistent forms.\cite{DJK1,DJK2}
However, a critical discussion of accuracy of this method in
multi-band situation and comparison with e.g. QMC have not yet
been carried out. It is the aim of this paper to fill this void,
and compare diagrammatic schemes with dynamical fluctuations
based on two-particle scatterings with finite-temperature quantum
Monte Carlo solution of the DMFT. To this end we use a
multiorbital Hubbard model with a simplified kinetic energy so
that we can focus our attention on the fundamental features of
the transition between weak and strong electron couplings in
multiorbital Hubbard models.

A typical form of the Hamiltonian used in dynamical extensions of
DFT schemes is
\begin{eqnarray} %
H^{\rm Hubb}& = &\sum_{{\bf R}\lambda,{\bf R'}\lambda'} t_{{\bf
R}\lambda,{\bf R'}\lambda'} \, a^{+}_{{\bf R}\lambda} \, a_{{\bf
R'}\lambda'}  \\ \nonumber%
 & + &\sum_{{\bf R},\lambda, \lambda'
\lambda'' \lambda'''} \langle {\bf R} \lambda, {\bf R}\lambda'|V|
{\bf R}\lambda'' {\bf R}\lambda''' \rangle \, a^{+}_{{\bf
R}\lambda} a^{+}_{{\bf
R}\lambda'} a_{{\bf R}\lambda'''} a_{{\bf R}\lambda''},\\\nonumber%
\label{mbhh}%
\end{eqnarray} %
where $\mathbf{R}$ are lattice site coordinates and
$\lambda=(l\sigma)$ are spin-orbital indices. The hopping term
$t_{{\bf R}\lambda,{\bf R'}\lambda'}$ is determined from {\em ab
initio} electronic structure calculations and will be replaced in
this comparison study with a model dispersion relation diagonal
in the spin-orbital indices. The electron interaction is usually
considered only between the $d$-electrons, since the effect of
the lower orbitals is assumed to be described quite well within
the standard DFT. We assume that the local interaction consists
only of direct and exchange terms. We approximate the interaction
operator with two parameters only: the Hubbard $U$ and the
exchange constant $J$. In homogeneous cases (without disorder) we
can neglect the lattice coordinate and represent the interaction
only with a quadruple of spin-orbital indices
\begin{equation}
\langle i \sigma j \sigma'|V| k \sigma l \sigma' \rangle \,
\approx  \delta_{ik} \delta_{jl} (1-
\delta_{ij}\delta_{\sigma\sigma'}) \, U
 + \delta_{il} \delta_{jk} (1 -
\delta_{ij})\delta_{\sigma\sigma'} \, J \, . \label{ujapprox}
\end{equation}
This representation can easily be further simplified to a
standard matrix in the spin-orbital indices
\begin{equation}
  \label{eq:int_simple}
  v_{\lambda\lambda'} =
 (1-\delta_{\lambda\lambda'})(U-J\delta_{\sigma\sigma'}) \ .
 \end{equation}
We use this representation of the electron interaction in our
many-body treatment of the multiorbital Hubbard
 model.

\section{Methods}

\subsection{Many-body dynamical fluctuations} \label{sec:FLEX}

The effects of the electron interaction on one-particle  states
are described by the self-energy $\Sigma_{\lambda}$. Dynamical
fluctuations are contained in the two-particle vertex function
$\Gamma_{\lambda\lambda'}$. We denote four-momenta
$k=(\mathbf{k},i\omega_n)$ and $q=(\mathbf{q},i\nu_m)$,  and use
the Schwinger-Dyson equation to relate the two-particle vertex
with the one-particle self-energy. With representation
(\ref{eq:int_simple}) for the electron interaction we can write
\begin{eqnarray}
  \label{eq:SDE}
  \Sigma_{\lambda}(k)& =& \sum_{\lambda'} \frac 1{\beta N} \sum_{k'}
  v_{\lambda\lambda'} G_{\lambda'}(k')\times \\\nonumber
  && \left[1 - \frac 1{\beta N}
    \sum_{q}G_{\lambda}(k-q) G_{\lambda'}(k'-q) \Gamma_{\lambda\lambda'}
    (k-q; q, k'-k) \right]\ .
\end{eqnarray}
The first term on the r.h.s. of Eq.~(\ref{eq:SDE}) is the  static
Hartree term expressing the self-energy in terms of densities.
This term in realistic calculations is normally part of the
static local density approximation fixing the static particle
densities. We hence suppress the Hartree term and use only the
vertex contribution to the self-energy as a generator of
dynamical fluctuations missing in the DFT.

The simplest approximation on the vertex function
$\Gamma_{\lambda\lambda'}$ is the bare interaction
$v_{\lambda\lambda'}$. Such an approximation corresponds to
second-order perturbation theory (SOPT). The vertex is momentum
independent. Even in more advanced approximations we will not use
the full momentum dependence of the vertex function. In our
treatment we resort only to multiple scatterings of two
quasiparticles (FLEX)~\cite{BS}. In this situation the
two-particle vertex depends on only a single bosonic
four-momentum $q$. This dependence enters the vertex function via
a two-particle bubble. When we deal with multiple electron-hole
scatterings the bubble is \begin{equation}
  \label{eq:bubble_eh}
  \Phi_{\lambda\lambda'}(q) = \frac 1{\beta N} \sum_{k} G_{\lambda}(k)
  G_{\lambda'} (k+q)\ .
\end{equation}
The self-energy due to dynamical electron-hole (multiple)
scatterings can then be represented as \begin{equation}
  \label{eq:eh-channel}
  \Sigma^{eh}_{\lambda}(k) = \sum_{\lambda'}\frac 1{\beta N}\sum_{q}
  v_{\lambda\lambda'} G_{\lambda'}(k+q)\Phi_{\lambda\lambda'}(q)
  \Gamma^{eh}_{\lambda\lambda'}(q)\ .
\end{equation}

\begin{figure}[h]
  \includegraphics[width=0.60\linewidth,angle=0]{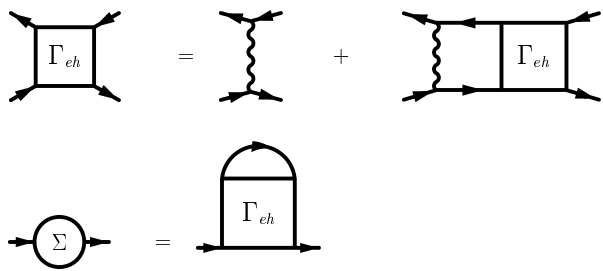}
\caption{Particle--hole (top row) vertex $\Gamma^{eh}$ which appears in the
FLEX approximation. Definition of the FLEX self--energy
constructed with the help of the  particle--hole  vertex
(bottom row).
  \label{fig:01}}
\end{figure}

In second-order perturbation theory
$\Gamma^{eh}_{\lambda\lambda'} = v_{\lambda\lambda'}$. When we
sum ladder electron-hole diagrams we obtain for the vertex
function the following representation \begin{equation}
  \label{eq:BS-eh}
   \Gamma^{eh}_{\lambda\lambda'}(q) = \frac{v_{\lambda\lambda'}}
   {1+v_{\lambda\lambda'}\Phi_{\lambda\lambda'}(q)}\ .
   \end{equation}
To visualize the above equations we show corresponding
diagrams for the vertex $\Gamma^{eh}$ of the particle--particle type shown in
the top row of Fig.~\ref{fig:01} and the corresponding contribution
to the self-energy is shown in the bottom row of Fig.~\ref{fig:01}.
Note that a small square in all diagrams (Fig. 1 -- 3) represents
the antisymmetrized pair interaction (cf. Ref. \cite{AGD}).

\begin{figure}[h]
  \includegraphics[width=0.60\linewidth,angle=0]{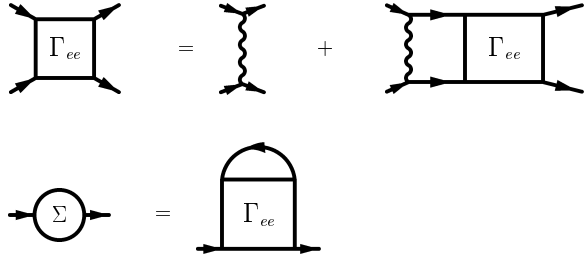}
\caption{Particle--particle (top row) vertex $\Gamma^{ee}$ which appears in
the FLEX approximation. Definition of the FLEX self--energy
constructed with the help of the particle--particle vertex
(bottom row).
  \label{fig:02}}
\end{figure}

Analogously we can construct an approximation with multiple
electron-electron scatterings where the self-energy can be
represented as
\begin{equation}
  \label{eq:ee-channel}
  \Sigma^{ee}_{\lambda}(k) = \sum_{\lambda'}\frac 1{\beta N}\sum_{q}
  v_{\lambda\lambda'} G_{\lambda'}(q-k)\Psi_{\lambda\lambda'}(q)
  \Gamma^{ee}_{\lambda\lambda'}(q)\ .
\end{equation}
Here we have to use a particle-particle bubble
\begin{equation}
  \label{eq:bubble_ee}
  \Psi_{\lambda\lambda'}(q) = \frac 1{\beta N} \sum_{k} G_{\lambda}(k)
  G_{\lambda'} (q-k)\ .
\end{equation}
The two-particle vertex $\Gamma^{ee}_{\lambda\lambda'}$ has the
same solution as the electron-hole scattering function,
Eq.~(\ref{eq:BS-eh}), where only we replace the bubble $\Phi$ with
$\Psi$. Corresponding set of diagrams for the vertex $\Gamma^{ee}$ and the
self-energy is presented in Fig.~\ref{fig:02}.

\begin{figure}[h]
 \includegraphics[width=0.60\linewidth,angle=0]{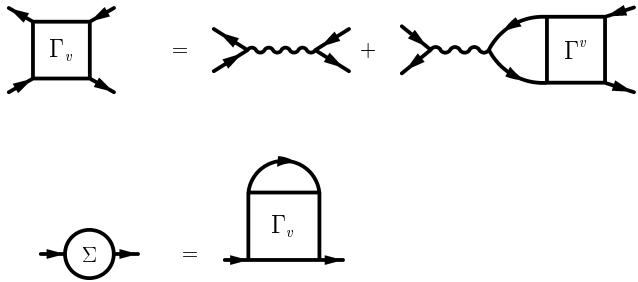}
\caption{Particle--hole vertex $\Gamma^v$ corresponding to the third
scattering channel is given in top row. Definition of the
corresponding self--energy is presented in the bottom row.
  \label{fig:03}}
\end{figure}

The third channel of two-particle scatterings is the interaction
channel where the electron interaction is screened by
electron-hole polarization bubbles. The dynamical self-energy due
to this renormalization  is then represented as
\begin{equation}
\label{eq:v-channel}
\Sigma^{v}_{\lambda}(k) =
\sum_{\lambda'}\frac 1{\beta N}\sum_{q}
  v_{\lambda\lambda'} G_{\lambda' }(k+q)\Phi_{\lambda\lambda'}(q)
  \Gamma^{v}_{\lambda\lambda'}(q),
\end{equation}
with
\begin{equation}
  \label{eq:BS-v}
   \Gamma^{v}_{\lambda\lambda'}(q) = v_{\lambda\lambda'} -
    \sum_{\lambda''} v_{\lambda\lambda''}\Phi_{\lambda''\lambda''}(q)
    \Gamma^{v}_{\lambda''\lambda'}(q)\ .
\end{equation}
Corresponding diagrams are presented in Fig.~\ref{fig:03}.
We can treat each channel independently or add all three channels
to assess the effect of dynamical fluctuations on the electron
self-energy. In the latter case, however, we have to subtract
twice the contribution from second order, since it is identical
in all three channels.

The idea of the DMFT is to neglect the momentum dependence of the
one-electron propagators in the contributions to the self-energy.
We hence use only the local parts of the one-electron
propagators, i.~e., $G_{\lambda}(\mathbf{k},i\omega_n)\to
N^{-1}\sum_{\mathbf{k}} G_{\lambda}(\mathbf{k},i\omega_n)=
G_{\lambda}(i\omega_n)$. Then, all the above formulas hold with
the replacements $k\to i\omega_n$, $q\to i\nu_m$ for fermionic
and bosonic momenta, respectively.

The advantage of analytic approaches with multiple two-particle
scatterings is the knowledge of the explicit analytic structure
of the self-energy. Hence all the above results can be explicitly
analytically continued to real frequencies using a
straightforward procedure. We explicitly mention only the result
for the two-particle electron-hole bubble
\begin{equation}  \label{eq:bubble_cont}
 \Phi_{\lambda\lambda'}(z) =
-\int_{-\infty}^{\infty}\frac{d\omega}{\pi}
 f(\omega-\mu) \left[G_{\lambda'}(\omega+z) {\rm Im}
G_{\lambda}(\omega_+) +
 G_{\lambda}(\omega-z){\rm  Im }G_{\lambda'}(\omega_+) \right]\ .
\end{equation} %

Up to now we have used the fully renormalized one-electron
propagators in the perturbation theory as demanded by
conservation laws. However, when we are interested in the
one-electron properties of the system, we can relax the demands
of thermodynamic consistence and replace the fully renormalized
propagator with a partially renormalized one

\begin{equation}
\label{eq:bath-GF} %
G^{-1}(z)\longrightarrow \mathcal{G}_{0}^{-1}(z) =
G^{-1}(z)+\Sigma(z) + \tilde\mu ,
\end{equation}
where  in the expression for $G_0$ we used  an additional shift of
the impurity level to satisfy Luttinger's theorem following
Ref.~\cite{Kajueter:1996:thesis}. This propagator goes over into
the bare propagator in the atomic limit and does not contain long
energy tails due to frequency convolutions in the definition of
the self-energy.

This partially non-self-consistent scheme resembles  IPT. Hence a
better description of the transition from weak to strong coupling
regimes at the one-particle level, including the metal-insulator
transition (MIT), is expected from the IPT than from the
conserving scheme with fully renormalized propagators, where the
metal-insulator transition at half filling is known to be missing.

\subsection{The quantum Monte Carlo}%
\label{sec:MC}

Among many methods used to solve the impurity problem we choose
the quantum Monte Carlo method~\cite{Hirsch:1986} to benchmark
FLEX as a potential candidate for impurity solver. There are well
known advantages and disadvantages of the QMC method and our
choice is spurred by the fact that despite being slower than
other methods, the QMC is well controlled, numerically exact
method.  As an input the QMC procedure gets Weiss function ${\cal
G}_{0}(\tau)$ and as an output it produces Green's function
$G(\tau )$. We remind the reader  major steps taken in the QMC
procedure. Usually one starts with an impurity effective action
$S$:
\begin{equation}
S_{\rm eff}= - \int^{\beta}_{0} d\tau d\tau' \sum_{\alpha}
c^{+}_{\alpha}(\tau) {{\cal G}_{0}}_{\alpha}^{-1}(\tau,\tau')
c_{\alpha}(\tau^{\prime}) +\frac{1}{2}\,\int^{\beta}_{0}d\tau\,
\sum_{\alpha ,\alpha '}
U_{\alpha\alpha'}n_{\alpha}(\tau)n_{\alpha'}(\tau),
\label{eq:simp}
\end{equation}
where $\{c,\ c^+\}$ are fermionic annihilation and
creation operators of the lattice problem, $\alpha =\{m,\sigma\}$.

The first what we should do with the action (\ref{eq:simp})  is
to discretize it in imaginary time with time step $\Delta \tau$
so that $\beta=L\Delta\tau$, and $L$ is the number of time
intervals: %
\begin{equation} S_{\rm eff} \rightarrow
\sum_{\alpha ,\tau \tau ^{\prime } }c_{\alpha }^{+}(\tau ) {{\cal
G}_{0}}_{\alpha }^{-1}(\tau ,\tau ^{\prime })c_{\alpha  }(\tau
^{\prime }) + \frac{1}{2}\sum_{\alpha ,\alpha '}U_{\alpha\alpha'}
n_{\alpha }(\tau)n_{\alpha '}(\tau). \label{eq:simdiscret}
\end{equation} The next step is to get rid of the interaction
term $U$ by substituting it by summation over Ising-like
auxiliary fields. The decoupling procedure is called the
Hubbard-Stratonovich
transformation~\cite{Hirsch:1983,Takegahara:1992}:
\begin{equation}
\exp \{  - \Delta \tau  \{U_{\alpha\alpha'} n_{\alpha}  n_{\alpha
'} - \frac{1}{2}(n_{\alpha}   + n_{\alpha '} )\} \}  =
\frac{1}{2}\sum\limits_{S_{\alpha \alpha '} =  \pm 1} {} \exp \{
\lambda_{\alpha \alpha '} S_{\alpha \alpha '} (n_\alpha   -
n_{\alpha '} )\}, %
\label{eq:HStransformation}
\end{equation}
where $ \cosh \lambda_{\alpha \alpha '}  = \exp (\frac{{\Delta
\tau U_{\alpha \alpha '} }}{2}) $ and $S_{\alpha \alpha
'}(\tau_l)$ are auxiliary Ising fields at each time slice.

In the one-band Anderson impurity model  we have only one
auxiliary Ising field $S(\tau_l)=\pm 1$ at {\sl each time slice},
whereas in the multiorbital case  number of auxiliary fields is
equal to the number of $\alpha, \alpha'$ pairs. Applying the
Hubbard-Stratonovich transformation at each time slice we bring
the action to a quadratic form with the partition function:
\begin{equation}  Z = {\rm Tr}_{\{S_{\alpha \alpha '}(\tau)\}}
\prod_{\alpha } \det G^{-1}_{\alpha ,\{S_{\alpha \alpha
'}(\tau)\} }\; , \end{equation} where the Green function in terms
of auxiliary fields $G^{-1}_\alpha $ reads
\begin{equation}
G^{-1}_{\alpha ,\{S_{\alpha \alpha '} \}}(\tau, \tau') = {{\cal
G}_0}^{-1}_{\alpha }(\tau, \tau')e^{V} -
(e^{V}-1)\delta_{\tau,\tau'}, \end{equation} with the interaction
matrix \begin{equation} V_{\tau}^{\alpha } = \sum\limits_{\alpha
'(\ne \alpha )} \lambda_{\alpha \alpha '}  {S_{\alpha \alpha '}
(} \tau ) \sigma_{\alpha \alpha '},
\end{equation}
where\\  $ \sigma_{\alpha \alpha '}  =  + 1 $ for $\alpha  < \alpha ' $\\
 $ \sigma_{\alpha \alpha '}  =  - 1 $ for   $\alpha  > \alpha '$.\\ %

Once the quadratic form is obtained one can apply Wick's theorem
at each time slice and perform the Gaussian integration in
Grassmann variables to get the full interacting Green function:
\begin{equation} \!\!\!\!\!\!\!\! G_{\alpha }(\tau,\tau') =
{\frac{{1}}{{Z}}}\, {\rm Tr}_{\{ S_{\alpha \alpha '} \} }
\nonumber  G_{\alpha ,\{ S_{\alpha \alpha
'}\}}(\tau,\tau')\prod_{\alpha '} \det G^{-1}_{\alpha
',\{S_{\alpha \alpha '}\}}. \label{eq:gfull}%
 \end{equation}

To evaluate summation in Eq.~(\ref{eq:gfull}) one  uses Monte
Carlo stochastic sampling. The product of determinants is
interpreted as the stochastic weight and auxiliary spin
configurations are generated by a Markov process with probability
proportional to their statistical weight. More rigorous
derivation can be find elsewhere ~\cite{GKKR},
\cite{Takegahara:1992}.

Since the QMC method produces results in imaginary  time axis
($G(\tau_m)$ with $\tau_m=m\Delta\tau$, $m=1...L$) and the DMFT
self-consistency equations make use of the frequency dependent
Green's functions and self-energies we must have an accurate
method to compute Fourier transforms from the time to the
frequency domain. This is done by representing the functions in
the time domain by a cubic splined functions which should go
through original points with condition of continuous second
derivatives imposed. Once we know cubic spline coefficients we
can compute the Fourier transformation of the splined functions
analytically.

\section{Results and Discussion}
\label{sec:results}

To compare analytic solutions with  numerical ones we employed a
simple model with two degenerate bands, i.e., four spin-orbitals
per site and assumed a semi-elliptic density of states (DOS),
$\rho_0(E)=\frac{2}{\pi} \sqrt{1-E^2}$. For simplicity we
resorted to the case $J=0$ and non-magnetic solutions. We tested
second-order perturbation theory together with multiple
scatterings from the electron-hole and electron-electron
interaction channels. Both types of self-consistency were
considered, i.~e., the full conserving and the partial one with
the bath function $\mathcal{G}_0$, defined in the IPT
Eq.~(\ref{eq:bath-GF}). All approximations were analytically
continued to real frequencies before being evaluated numerically.
Maximum entropy method was used for analytical continuation to
real axis. We tested maximum entropy method against other
high-frequency more reliable methods like non-crossing
approximation and one-crossing approximation  and found
satisfactory agreement understandable within limitations of all
used methods. A simple iteration procedure with a suitably chosen
mixing of the old and new self-energy led to well converged
results for moderate values of the pair interaction $U$. The
details of the numerical implementation of multiorbital FLEX-type
approximations were described elsewhere \cite{DJK1, DJK2, DJK3,
Katsnelson:1999, Katsnelson:2000}.

\begin{figure}[h]
\begin{center}
\includegraphics[scale=0.5]{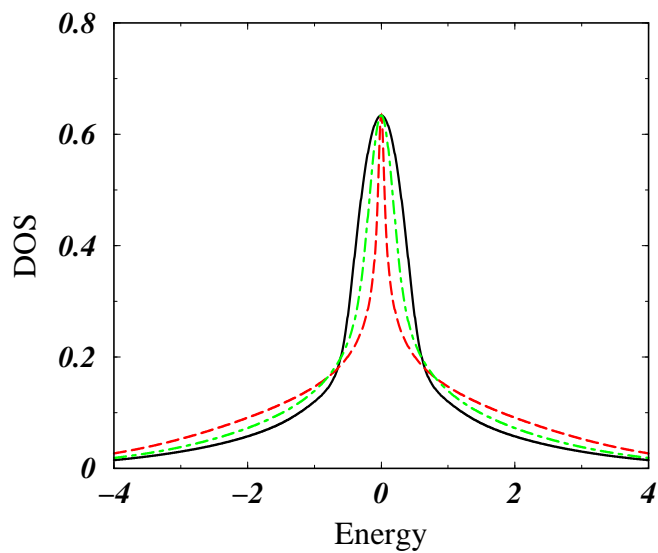}
\end{center}%
\caption{\label{Fig4}Densities of states for $U=2$, $n=2$
(half-filled case) calculated within the second-order
perturbation theory (dot-dashed line), electron-electron TMA
(solid line), and electron-hole TMA  (dashed line). The
one-particle propagators are fully renormalized in the
self-consistent calculations
($\Sigma=\Sigma[G]$).} %
\end{figure}

First, we compared the analytical approximations among themselves
for an intermediate value of the interaction strength, $U=2$. All
energies are given in units of the half-bandwidth of the
non-interacting DOS, $D=1$. We also wrote two different FLEX
programs on real and imaginary axes and used 9000 and 4096 points
on corresponding frequency domain  with large enough frequency
cut-off ($10D$) for temperature $T=1/16~D$. Using the conserving
full self-consistence  we found that results of SOPT and the
electron-electron $T$-matrix approximation TMA are quite close to
each other. The addition of the electron-hole TMA channel changes
the result significantly. The quasiparticle width gets
substantially reduced as shown in figure~\ref{Fig4}. This strong
band narrowing is unphysical. We did not include the third
channel, screening by the electron-hole polarization bubbles,
explicitly to this figure, since it contributes quantitatively
and qualitatively similarly to the electron-hole ladder. The
reason why the electron-hole scatterings do not improve upon
second order is quite clear. At the mean-field level an
instability towards the local moment state with non-zero $\langle
n_\uparrow\rangle -\langle n_\downarrow\rangle $ appears at some
critical interaction. This instability is totally artificial
because the conduction electrons screen the local moment (Kondo
effect). In the non-self-consistent FLEX approach including both
electron-electron and electron-hole TMA, this instability also
occurs when the denominator in the above equation contains a pole
at zero frequency, for example for two-band Hubbard model with
half bandwidth equal to one the instability happens at $U=0.7$,
for temperature $T=1/16$. So the presence of this artificial
instability will push the results obtained by non-self-consistent
FLEX approach including electron-hole TMA  away from the right
answer. For the self-consistent FLEX approach including
electron-hole TMA, the instability happens for $U$ slightly larger
than two for the same model, indicating tendency to a phase
transition from the paramagnetic to a ferromagnetic solution in
the lattice case what is unphysical for the impurity problem.
Thus, this unphysical divergence overestimates the contribution
of the electron-hole scatterings and the result worsens with
increasing the interaction strength. To improve this one can use
Kanamori's ~\cite{Kanamori:1963} observation that the
particle-hole bubbles should interact not with the bare
interaction but with an effective one screened by the
particle-particle ladder i.e. replace $U$ by $U_{eff}$ equal to
electron-electron $T$-matrix taken at zero frequency.

\begin{figure}[h]
\begin{center}
\includegraphics[scale=0.5]{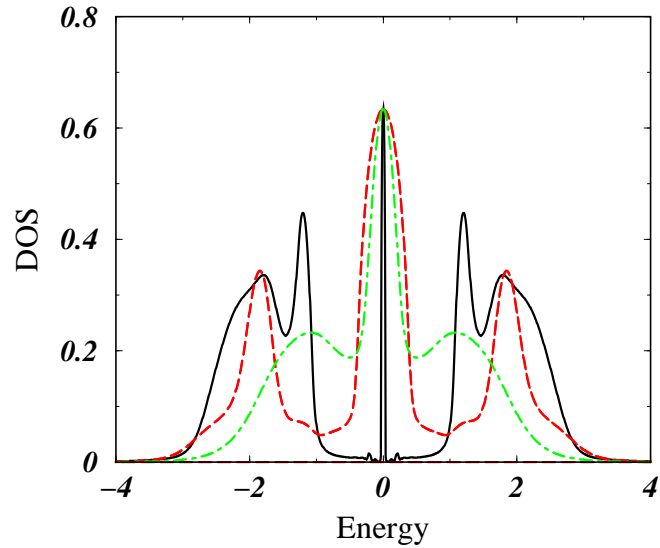}
\end{center}%
\caption{\label{Fig5}Densities of states for $U=2$, $n=2$
(half-filled case) calculated within the second-order
perturbation theory (solid line), electron-electron TMA (dashed
line), and electron-hole TMA (dot-dashed line). The one-particle
propagators are partially renormalized by using the iterated
perturbation method
($\Sigma=\Sigma[\mathcal{G}_0]$).} %
\end{figure}

As it is known from the single-band situation, neither of the
FLEX-type approximations is able to trace the emergence of
satellite Hubbard bands and the metal-insulator transition. The
high-energy spectrum is shapeless and broadened due to energy
convolutions. This situation changes dramatically if we keep only
the topological self-consistence of the one-particle propagators
obtained from the IPT, see figure~\ref{Fig5}. We can see that
second order and the electron-electron TMA produces the Hubbard
satellite bands in their correct positions. The SOPT scatterings
add an additional internal structure to the satellite bands.
While the satellite bands gain more weight than they actually
have at this interaction strength, the central quasiparticle peak
weight is strongly reduced in non-self-consistent or partly
self-consistent approximations. Too much quasiparticle weight is
transferred to the Hubbard bands.

In the electron-hole scattering
channel we get slight changes in the positions of the Hubbard
bands, and slight decrease of the quasiparticle width. Even for
rather weak interactions $U < 1$ one observes nearly total
disappearance of the quasiparticle peak when including the
electron-hole scattering channel. Due to this unphysical behavior
at moderate electron coupling and the instability towards  the
local state formation we leave the electron-hole scattering
channel out from further considerations.
There is no way at the
FLEX level to improve upon the results from the electron-electron
scattering channel. The best we could do is to add contributions
from all three distinct channels. But we had to subtract second
order contribution to the self-energy which dominates at moderate
and intermediate couplings so that we obtain negative density of
states at the Fermi energy. The only real improvements can be
reached by a new self-consistent coupling of the
electron-electron and electron-hole channels of the parquet type.
Since we are not yet able to implement parquet-type
self-consistence in multiorbital models and have to stay within
FLEX, we compare only results of analytical approximations based
on second order perturbation theory and electron-electron
$T$-matrix against QMC data.

\begin{figure}[h]
\begin{center}
\includegraphics[scale=0.5]{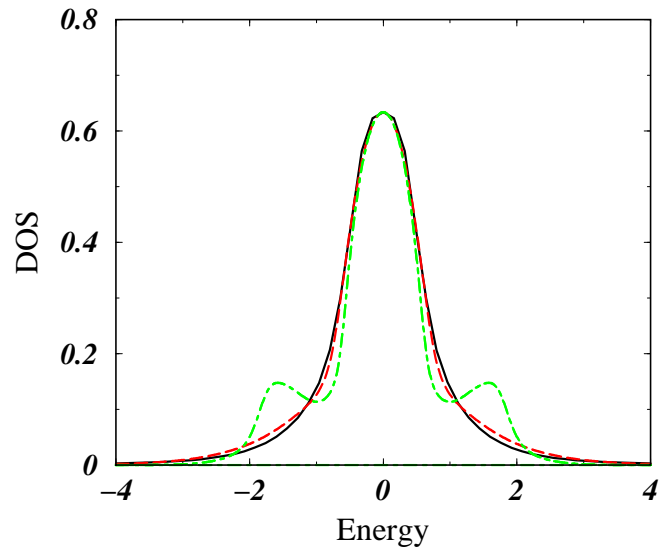}
\end{center}%
\caption{\label{Fig6}Densities of states for $U=1.0$, $n=2$
(half-filled case) calculated within the electron-electron TMA
using the full renormalization of the one-particle propagators
(dashed line) and partial renormalization using the IPT
(dot-dashed line) compared
with the result of quantum Monte Carlo method (solid line).}%
\end{figure}

\begin{figure}[h]
\begin{center}
\includegraphics[scale=0.5]{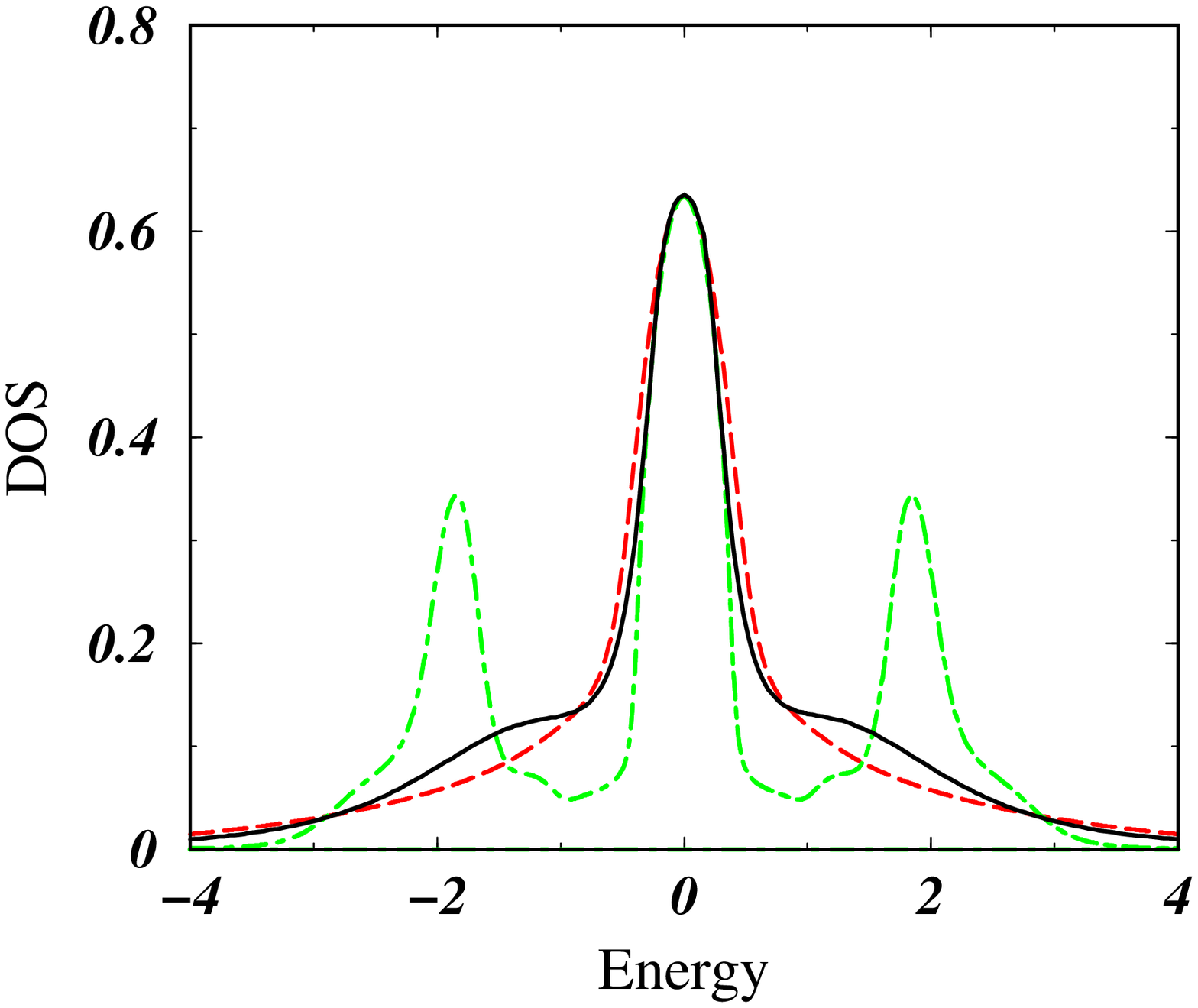}
\end{center}%
\caption{\label{Fig7}Densities of states for $U=2.0$, $n=2$
(half-filled case) calculated within the electron-electron TMA
using the full renormalization of the one-particle propagators
(dashed line) and partial renormalization using the IPT
(dot-dashed line) compared
with the result of quantum Monte Carlo method (solid line).}%
\end{figure}

To compare the analytic FLEX results with quantum Monte Carlo
simulations, we use both versions of the self-consistency. In
figure~\ref{Fig6} the density of states at $U=1$ calculated within
the electron-electron $T$-matrix approximation is shown along
with the one deduced from the QMC. As one can see from the plot,
the conserving TMA fits better the quasiparticle peak than the
TMA with the IPT bath Green function $\mathcal{G}_0$. In
addition, the IPT TMA seems to overemphasize the role of the
satellite peaks at weak coupling as there are no satellites in
the QMC solution for this coupling strength. If we increase the
interaction to $U=2$ (see figure~\ref{Fig7}) and compare the same
curves as in the previous plot, we notice that the conserving TMA
still very well reproduces the quasiparticle peak, whereas the
IPT TMA keeps the tendency to reducing the central peak in favor
of the satellites. At this coupling the satellites are formed
also in the QMC solution but not as strongly as the IPT TMA
predicts. Notice that positions of the satellites reproduced by
the IPT solution are rather close to the ones coming from QMC.

\begin{figure}[h]
\begin{center}
\includegraphics[scale=0.35]{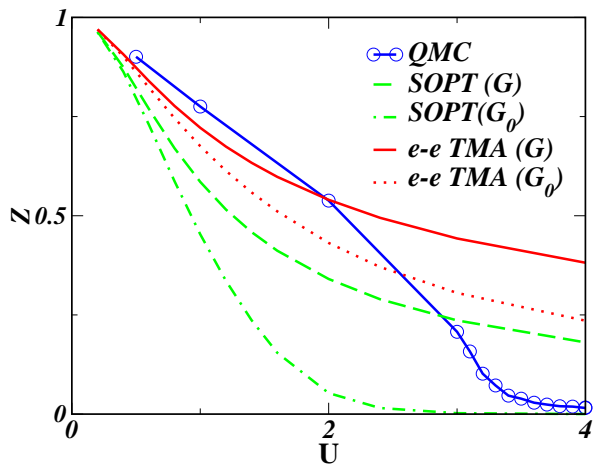}
\end{center}%
\caption{\label{Fig8}Dependence of the quasiparticle residue $Z$
on Coulomb repulsion $U$ calculated within the electron-electron
TMA (solid and dotted with crosses lines) and the second-order
perturbation theory (dashed and dot-dashed  lines) using the full
renormalization of the one-particle propagators and partial
renormalization using the IPT correspondingly. QMC results are
plotted by solid line with circle symbols.} %
\end{figure}

As a summary of the FLEX  for the symmetric case of the multiband
Hubbard model we present dependence of the quasiparticle residue,
$Z$, on the interaction strength, $U$. It is clear that the
closer $Z(U)$ curve for a particular approximation to the QMC data
the better the approximation works in reproducing the
quasiparticle properties of the system, including the central
peak weight and width. In figure~\ref{Fig8} we presented the
following curves: QMC data are plotted by solid curve with open
circle symbols, results of the second order perturbation theory
in the cases of fully renormalized one-particle propagators in
the self-consistent calculations ($\Sigma=\Sigma[G]$) and
partially renormalized by using the iterated perturbation method
($\Sigma=\Sigma[\mathcal{G}_0]$) are given by dashed and
dot-dashed lines correspondingly. Concentrating on small and
intermediate values of interaction, we can see that in the first
case of the full renormalization of the propagators the situation
is significantly better as the dashed curve goes much closer to
the QMC one than the dot-dashed curve. The electron-electron TMA
gives even better results than those obtained from SOPT. Both the
electron-electron TMA curves, for the case of the full
self-consistency ($\Sigma=\Sigma[G]$) plotted by solid line and
the partial one ($\Sigma=\Sigma[\mathcal{G}_0]$), plotted by
dotted line, lay much closer to the QMC data than the SOPT curves
(for interactions smaller or equal to $U=2$). From the two
electron-electron TMA curves we make a decision in favor of the
one obtained using the fully renormalized one-particle
propagators in the self-consistency procedure. This result
differs from the findings in the case of the one-band Hubbard
model, where partially renormalized propagators gave better
agreement with Monte Carlo results. Note that the FLEX-type
approximations follow the Monte Carlo quasiparticle weight only
for weak and moderate interaction strengths, $U\le 2$. All
approximations except for IPT go wrong near and beyond the
expected metal-insulator transition.

\begin{figure}[h]
\begin{center}
\includegraphics[scale=0.5]{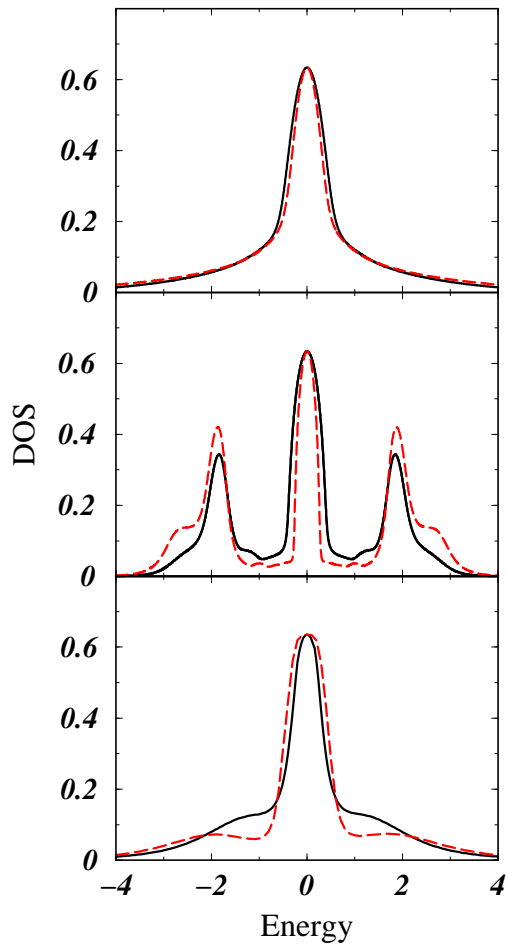}
\end{center}%
\caption{\label{Fig9}Densities of states for $U=2.0$, $n=2$ in the
two-band model (solid lines) and $n=3$ in the three-band model
(dashed lines)  (both cases correspond to half-filled case)
calculated within the electron-electron TMA using the full
renormalization of the one-particle propagators (upper panel) and
partial renormalization using the IPT (middle panel) compared
with the result of quantum Monte Carlo method (lower panel).}
\end{figure}

In  figure~\ref{Fig9} we study dependence of DOS, calculated by
different methods, on orbital degeneracy. We calculated DOS within
two- and three-band Hubbard models in the half-filled case using
QMC as the tester against two best FLEX approximations which we
found, namely, electron-electron TMA with the full
self-consistency ($\Sigma=\Sigma[G]$) and the partial one
($\Sigma=\Sigma[\mathcal{G}_0]$). One can almost immediately
notice the main difference between the QMC and FLEX results. The
width of quasiparticle (QP) peak in  QMC  calculation is slightly
increased with the degeneracy. This is expectable as while the
degeneracy grows the critical $U$ also increases what brings the
curve $Z$ versus $U$ dependence (see figure~\ref{Fig8})  to go
higher for the larger degeneracy at any particular repulsion.  As
an example the quasiparticle residue against repulsion at $U=2$
for degeneracy $N=6$ will be above the curve in figure ~\ref{Fig8}
corresponding to degeneracy $N=4$. FLEX results show opposite
dependence: the QP width decreases with increasing degeneracy. We
should also notice that this wrong tendency is stronger for the
partial (IPT) self-consistency. The reason for such kind of
behavior can lay in a limited set of diagrams treated in FLEX
resulting in unsufficient screening for higher degeneracy.

\begin{figure}[h]
\begin{center}
\includegraphics[scale=0.5]{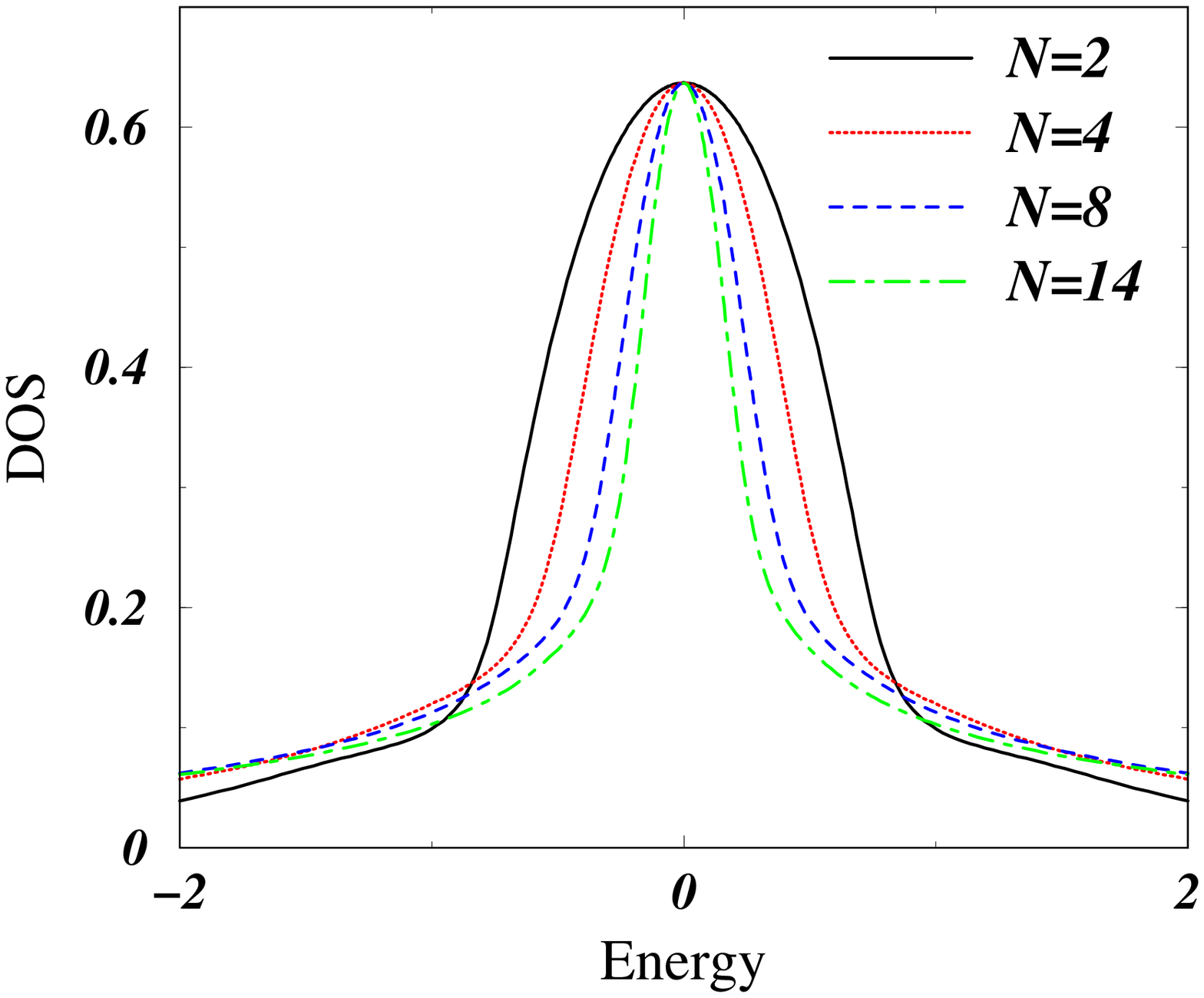}
\end{center}%
\caption{\label{Fig10}Densities of states for $U=2.0$ ($n$ is kept
at the half-filling) calculated within the electron-electron TMA
using the full renormalization of the one-particle propagators for
different degeneracy (see the legend). }%
\end{figure}

The self-consistent FLEX scheme was found  to reproduce quite
well the central quasiparticle peak for two and three bands at
weak and medium interaction strength. It is, however, important
to stress that this cannot persist to very large degeneracy. It
has been shown~\cite{Florens:2002} that in the exact solution of
the DMFT equations for large $N$ the critical $U$ where the Mott
transition takes place at zero temperature, scales linearly with
$N$. This implies that {\it for fixed $U$} the quasiparticle
residue, which has  an approximate expression, $ Z= 1- U/U_{c2}$
increases and eventually approaches unity with increasing orbital
degeneracy. This remarkable screening effect in the multiorbital
degenerate Hubbard model is  not captured by the FLEX approach
which displays the opposite trend, as can be seen in
figure~\ref{Fig10}.

\begin{figure}[h]
\begin{center}
\includegraphics[scale=0.5]{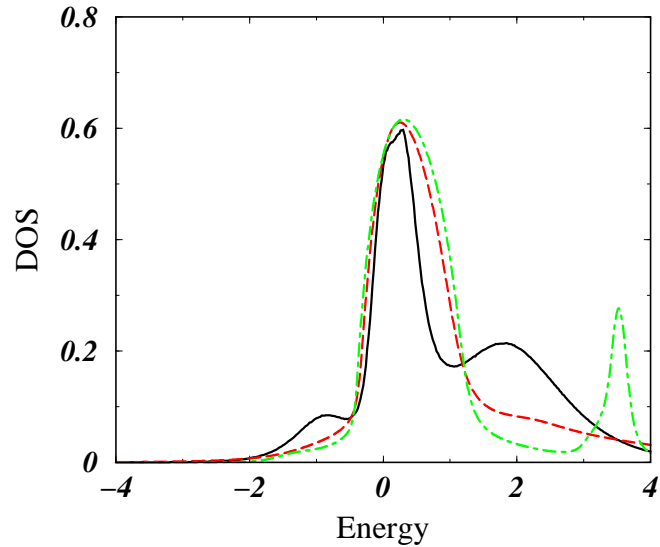}
\end{center}%
\caption{\label{Fig11} Densities of states for $U=2.0$, $n=0.8$
(partially filled band) calculated within the electron-electron
TMA using the full renormalization of the one-particle
propagators (dashed line) and partial renormalization using the
IPT (dot-dashed line) compared with the result of quantum Monte
Carlo method (solid
line).}%
\end{figure}

When we move off the electron-hole symmetric case, the situation
changes. The quasiparticle peak is no longer so strongly
suppressed in the electron-hole scattering channels. Moreover,
the conserving TMA  starts to develop a satellite peak.
Figure~\ref{Fig11} demonstrates this trend for $n=0.8$. We
perceive almost no difference in the form of the central peak in
the two different self-consistent versions of the multiple
electron-electron scatterings. The IPT version retains its
tendency to overemphasize the width of the satellite peaks and
produces an incorrect position of the upper Hubbard band. The
conserving TMA shows a shoulder behavior where the QMC displays
the hole satellite. The less pronounced electron peak cannot be
traced neither in the IPT TMA nor in the conserving TMA.

\begin{figure}[h]
\begin{center}
\includegraphics[scale=0.35,angle=0]{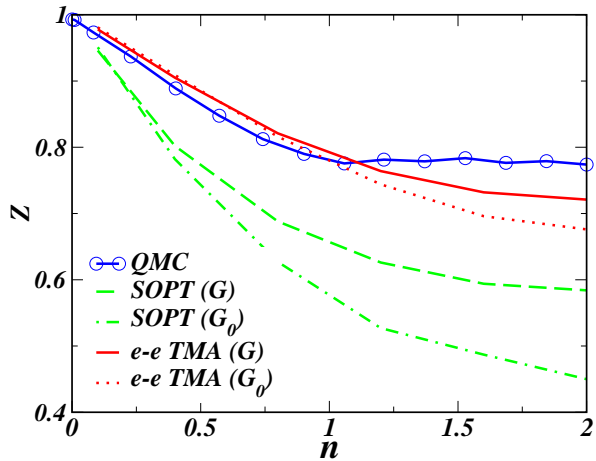}
\end{center}
\caption{\label{Fig12} Dependence of the quasiparticle residue $Z$
on filling $n$ for  $U=1$ calculated within the electron-electron
TMA (solid and dotted with crosses lines) and the second-order
perturbation theory (dashed and dot-dashed lines) using the full
renormalization  of the one-particle propagators and partial
renormalization using the IPT correspondingly. QMC results are
plotted by solid line with circle symbols.}
\end{figure}

To explore a wider range of parameters away from half-filling we
plot the quasiparticle residue dependence on filling in
figure~\ref{Fig12}. With the increment of filling $n$, the
scattering rate is increased as number of particles which can
scatter a particular electron in the system grows. Therefore it
is natural to expect the reduction of quasiparticle residue as
the function of $n$ at least for filling $n<1$. As one can see
from the plot, the relative positions of the curves are not
changed in comparison with figure~\ref{Fig8} as well as the
conclusion that the electron-electron TMA fits best the QMC data.
In addition, one can conclude that for densities $n<1$ both
methods (the self-consistent and the non-self-consistent
electron-electron TMA work very well describing QMC data, while
for $n>1$ we observe a deviation of FLEX from QMC data indicating
necessity to take into account other scattering channels. The
minima in $Z(n)$ at $n=1$ and $2$ observed in QMC data get even
more pronounced with increasing  $U$ and indicate the trace of
the metal-insulator transition which appears for the two-band
Hubbard model at $U\approx 3.5$ as one can see from
figure~\ref{Fig8}. As we mentioned above, the FLEX fails to
reproduce the MIT which is also reflected in an almost monotonic
decrease of the quasiparticle weight for all densities including
interval between $n=1$ and $n=2$.

{ %
We mention in passing that the one-band situation is exceptional
because the exact evaluation of the  one-particle Greens function
shows that interaction is not efficiently screened. Therefore,
the non-self-consistent version of SOPT theory works well.
Furthermore, in the one-band case, the SOPT form combined with
the DMFT self-consistency condition produces the correct atomic
limit \cite{GKKR}, this is not the case in the degenerate
situation. In the multiorbital case, the effective interaction is
more screened with increasing degeneracy. The self-consistent
version of FLEX, attempts to capture this effect, and in fact it
does it slightly better than the self-consistent SOPT.  The
non-self-consistent version of SOPT is worse because it does not
screen the interaction at all.
}

\section{Conclusion}
\label{sec:conclusion}

In summary,  we compared multiband implementation of the FLEX
procedure within the framework of the two-band model with
semi-elliptic non-interacting DOS against the quantum Monte Carlo
method. Obtained results indicate the best agreement with the QMC
data is reached for the electron-electron TMA and with fully
renormalized one-particle propagators, $\Sigma=\Sigma[G]$. This
conclusion is restricted to small energies within the
quasiparticle peak and to the region of moderate Coulomb
interaction, ($U_{} \le 2)$. Our finding is somewhat different
from that for the one-band Hubbard model, where the best match
for the same interaction strengths was found for the one-particle
propagators partially renormalized by using the iterated
perturbation method ($\Sigma=\Sigma[\mathcal{G}_0]$). In the
multiband situation for not high band degeneracy and small to
intermediate $U$ the self-consistent version describes the
quasiparticle features in good agreement with QMC. On the other
hand, there is no hint in the self-consistent calculation of an
incipient Mott transition, as in the one-band
case~\cite{Menge:1991}, while the non-self-consistent scheme
clearly gives a hint that a Mott transition will take place,
albeit at a very small unphysical value of $U$, due to an
improper screening of this interaction.

Unsufficient screening due to a limited set of diagrams treated
in FLEX results in a quasiparticle weight underestimation with
increasing degeneracy in the system.  For non-symmetric case,
FLEX has tendency to overestimate the QP width and underestimate
the Hubbard bands weight given their correct position in the case
of conserving TMA.
{ %
Our results, found for an impurity solver used within the DMFT,
have wider range of validity as they are naturally applied to the
multiorbital Anderson model as well.
}

\section*{Acknowledgments}
The research was carried out by V.D., V.J., and J.K. within the
project AVOZ1-010-914 of the Academy of Sciences of the Czech
Republic and supported in part by a grant A1010203 of the Grant
Agency of the Academy of Sciences of the Czech Republic. We
(V.O., X.D., K.H.,  and G.K.) acknowledge support from a grant
NSF-INT9907893 U.S.-Czech Materials Research on Many-Body
Correlations in Calculations of Realistic Electronic Structure of
Solids. We would like to thank A. Lichtenstein for extensive and
fruitful discussions of the works of FLEX. We also would like
acknowledge a warm hospitality extended to four of us  (V.J,
V.O., X.D. and G.K.) during our stay at Kavli Institute for
Theoretical Physics during the workshop ``Realistic Theories of
Correlated Electron Materials" where part of this work has been
carried out. The work was supported by the NSF grant DMR-0096462.


\vspace{1cm}







\begin{thebibliography}{99}

\bibitem{HK} Hohenberg P and Kohn  W 1964 {\it Phys. Rev.} {\bf 136} B864%

\bibitem{WK} Kohn W {\it Rev. Mod. Phys.} 1999 {\bf 71} 1253

\bibitem{GKKR} Georges A, Kotliar G, Krauth W and Rozenberg M
{\it Rev. Mod. Phys.} 1996 {\bf 68} 13

\bibitem{TDKSW} Turek I, Drchal V, Kudrnovsk\'y J, \v{S}ob M and
Weinberger P 1997 {\it Electronic Structure of Disordered Alloys,
Surfaces and Interfaces} (Boston-London-Dordrecht: Kluwer
Academic/Plenum Publishers)

\bibitem{Kajueter:1996} Kajueter H and  Kotliar G 1996
{\it Phys. Rev. Lett.} {\bf 77} 131-134

\bibitem{DJK1} Drchal V, Jani\v{s} V and  Kudrnovsk\'y J
1999 in {\it Electron  Correlations and Materials Properties}\
edited by A. Gonis et al. (New York: Kluwer Academic/Plenum
Publishers) p.~273

\bibitem{DJK2} Drchal V, Jani\v{s}  V and  Kudrnovsk\'y J %
1999 {\it Phys. Rev. } {\bf B 60} 15664

\bibitem{BS}
Bickers N E and Scalapino D J 1989  {\it Ann. Phys. (N.Y.)} {\bf 193} 206 %

\bibitem{AGD}
 Abrikosov A A, Gorkov L P and Dzyaloshinski I E 1975 {\it Methods of Quantum
 Field Theory in Statistical Physics} (New York: Dover)

\bibitem{Kajueter:1996:thesis}
 Kajueter H  1996 {\it Ph.D. thesis } (New
Brunswick   New Jersey: Rutgers University Graduate School)

\bibitem{Hirsch:1986}
Hirsch J E and Fye R M 1986 {\it Phys. Rev. Lett.} {\bf 56} 2521%

\bibitem{Hirsch:1983}  Hirsch J E 1983 {\it Phys. Rev. }  {\bf B 28} 4059 %

\bibitem{Takegahara:1992}  Takegahara K 1992 {\it J. Phys. Soc. Jpn.} {\bf 62} 1736%

\bibitem{DJK3}  Drchal V,  Jani\v{s} V and  Kudrnovsk\'y J 2003
in {\it Electron  Correlations and Materials Properties 2} edited
by A. Gonis et al. (New York: Kluwer Academic/Plenum Publishers)
p.~341.

\bibitem{Katsnelson:1999}
Katsnelson M I  and  Lichtenstein A I  1999 {\it J. Phys.: Cond. Matt.}  {\bf 11} 1037 %

\bibitem{Katsnelson:2000} Katsnelson M I and Lichtenstein A I 2000 %
{\it Phys. Rev. } {\bf B 61} 8906

\bibitem{Kanamori:1963} Kanamori J 1963 {\it Prog. Theo. Phys.} {\bf 30}
275

\bibitem{Florens:2002} Florens S, Georges A, Kotliar G and Parcollet O 2002 %
{\it Phys. Rev. } {\bf B 66} 205102

\bibitem{Menge:1991} Menge B  and M\"uller-Hartmann E 1991 {\it Z. Phys.~} {\bf B 82} 237

\end{thebibliography}
\end{document}